\documentclass[prd,showpacs,twocolumn]{revtex4}
\usepackage{amsmath}
\usepackage{amssymb}
\usepackage{epsfig}
\usepackage{graphicx}
\usepackage{array}
\usepackage{multirow}
\usepackage{color}
\usepackage{txfonts}
\usepackage[colorlinks, citecolor=blue, linkcolor=red, urlcolor=blue]{hyperref}
\allowdisplaybreaks[4]
\begin{document}
\renewcommand{\arraystretch}{1.5}
\newcommand{\psl}{ p \hspace{-1.8truemm}/ }
\newcommand{\nsl}{ n \hspace{-2.2truemm}/ }
\newcommand{\vsl}{ v \hspace{-2.2truemm}/ }
\newcommand{\epsl}{\epsilon \hspace{-1.8truemm}/\,  }
\newcommand{\tf}{\textbf}
\title{Effects from Hadronic Structure of Photon on $B\to\phi\gamma$ and $B_s\to(\rho^0,\omega)\gamma$ Decays}
\author{Yun Li$^1$}
\author{Zhi-Tian Zou$^2$}
\author{Yue-Long Shen$^3$}
\author{Ying Li$^{2}$} \email{liying@ytu.edu.cn}
\affiliation
{
$1$ Shaanxi Collaborative Innovation Center of Industrial Auxiliary Chemistry $\&$ Technology,
Shaanxi University of Science $\&$ Technology, Xi’an 710021, China\\
$2$ Department of Physics, Yantai University, Yantai 264005,China\\
$3$ College of Physics and Photoelectric Engineering, Ocean University of China,
Qingdao 266100, China}
\begin{abstract}
Using the perturbative QCD approach, we studied the effects from hadronic structure of photon on the pure annihilation rediative decays $B\to\phi\gamma$ and $B_s\to(\rho^0,\omega)\gamma$. These decays have small branching fractions due to the power suppression by the $\Lambda/m_B$, which make them very sensitive to the next-leading power corrections. The quark components and the related two-particle distribution amplitudes of a final state photon are introduced. The branching fractions can be enhanced remarkably by the factorizable and nonfactorizable emission diagrams. The branching fraction of $B\to \phi\gamma$ even increases by about 40 times, and those of $B_s \to \rho^0\gamma$ and $B_s \to \omega\gamma$ are at the order of ${\cal O}(10^{-10})$.  We also note that the ratio of branching fractions of $B_s \to \rho^0\gamma$ and $B_s \to \omega\gamma$ is very sensitive to the effects from hadronic structure of photon. All above results could be tested in future.
\end{abstract}
\maketitle
\section{Introduction}\label{sec:introduction}
The $B$ meson system, being a bound state that consists of a $b$ quark and a light antiquark, provides an ideal laboratory for precise study of the Standard Model (SM) of particle physics, and thus facilitates the search for new physics (NP). Because the $b$ quark mass is much larger than the typical scale of the strong interaction, the otherwise troublesome long-distance strong interactions are generally less important and are under better control than in other lighter meson systems. Radiative decays $B \to V \gamma$ are of particular interest in this respect. For example, the isospin-asymmetry parameter $\Delta^{\pm0}=\Gamma(B^{\pm}\to\rho^{\pm}\gamma)/[2\Gamma(B^0(\bar{B}^0)\to \rho^0\gamma)]-1$ and the direct $CP$ asymmetry parameter $\mathcal{A}_{CP}=[\mathcal{B}(B^-\to\rho^-\gamma)- \mathcal{B}(B^+\to\rho^+ \gamma)]/[\mathcal{B}(B^-\to\rho^-\gamma) +\mathcal{B}(B^+\to\rho^+\gamma)]$ allow us to extract the angle $\alpha$ of the Cobibbo-Kobayashi-Maskawa (CKM) unitary triangle \cite{Ali:2001ez}. The radiative $B\to K^*\gamma$ decays are usually viewed as probes of the NP \cite{Belle-II:2018jsg, LHCb:2018roe}, because they are induced by the flavor-changing-neutral-current $b\to s\gamma$ that only occurs by loops. Therefore, it is meaningful to improve the theoretical predictions of these radiative decays, so as to match with the coming accurate measurements in the ongoing LHCb and Belle II experiments.

In the past few years, many efforts \cite{Beneke:2001at, Bosch:2001gv,Beneke:2004dp,Bosch:2004nd,Lu:2005yz} have been devoted to improve the theoretical predictions of these exclusive radiative decays by including the various corrections, such as next-to-leading order in the strong coupling $\alpha_s$, the non-factorizable corrections, and the charm loop contributions. Each correction may affect the experimental observables, such as the branching fractions, $CP$ asymmetries and the forward-backward asymmetries. Some special radiative decays such as $B\to\phi\gamma$ and $B_s\to(\rho^0, \omega) \gamma$, where the quarks in the final state are different from ones in the initial $B$ meson, are called pure annihilation radiative decays. In particular, the annihilation diagrams are viewed as power suppressed by the $\Lambda/m_B$ with a typical hadronic scale $\Lambda$, therefore the branching fractions are estimated to be small. For the decay $B^0\to\phi\gamma$ which is induced by the penguin operators, its decay amplitude can be factorized into the simple matrix $\langle \phi|\bar s\gamma_\mu s|0\rangle$ and the transition form factor $\langle \gamma|\bar{d}\gamma_{\mu}(1-\gamma_5)b|B^0\rangle$ in the naive factorization. Its branching fraction was estimated to be at the order of ${\cal O}(10^{-13})$, and it would be ${\cal O}(10^{-12})$ \cite{Li:2003kz} by adding the QCD corrections. Within the perturbative QCD (PQCD) approach, its branching fraction had been predicted to be $1.6 \times10^{-12}$ \cite{Li:2006xe} by one of us (Li). These predictions are still much smaller than the current available upper limit $1.0\times10^{-7}$ \cite{Zyla:2020zbs}. As a decay induced only by the penguin operators, the branching fraction might be enhanced remarkably by the effects of NP, such as in the $R$-parity violation supersymmetry model \cite{Li:2003kz} and the non-universal $Z^\prime$ model \cite{Hua:2010we}. From this respect, before judging the effects of NP, we should calculate the branching fractions in SM as precise as possible.

As we know, besides the high order corrections of $\alpha_s$, the power corrections are also important for the finite $b$ quark mass. In the $B$ meson decays, there are many kinds of power corrections, such as that from the high-twist distribution amplitudes of $B$ meson \cite{Beneke:2011nf, Braun:2012kp, Shen:2020hsp}, the high-twist light-cone distribution amplitudes (LCDAs) of the final states meson \cite{Wang:2016qii, Beneke:2018wjp} and from the hadronic structure of the photon (HSP)\cite{Wang:2018wfj, Shen:2018abs} in radiative decays. In the framework of PQCD, the power corrections for $B\to\gamma l \nu$ decay were also investigated \cite{Shen:2018abs}, which indicated that both the contribution from the high-twist $B$ meson wave functions and the HSP can change the leading power result by about $20\%$. In view of the power counting rules, the corrections from high-twist LCDAs of light mesons are no more than $10\%$ for nonleptonic $B$ decays \cite{Kurimoto:2001zj}.  For the radiative nonleptonic $B$ decays that only occur through annihilation diagrams at leading power, when HSP is considered, the contributions from emission diagrams are involved, and the branching ratio might be enhanced. So, the pure annihilation decays are perhaps sensitive to the power corrections. Motivated by this, we shall employ the PQCD approach to investigate the corrections from HSP to the $B\to\phi\gamma$ and $B_s\to (\rho, \omega) \gamma$, so as to improve the precision of the theoretical prediction.

The outline this paper is as follows. In Sec.~\ref{sec:framework} we briefly review the theoretical background and summarize the expressions for the $B\to\phi\gamma$ and $B_s\to (\rho, \omega)\gamma$ amplitudes. In Sec.~\ref{sec:results}, the numerical results and  discusses are presented. We conclude in Sec.~\ref{sec:summay}.
\section{Framework and Inputs}\label{sec:framework}

It is well known that, when facing to the annihilation type contributions, in the collinear factorization such as QCD factorization, the singularities destroy the perturbative calculations, and then parametrization is adopted to evaluate this type contribution, which leads to the descent of the predictive ability. Using the PQCD approach \cite{Keum:2000ph,Lu:2000em} that is based on the $k_T$ factorization, one can calculate the annihilation diagrams perturbatively, because the end-point singularities are smeared by keeping the intrinsic transverse momenta of the inner quarks. The pure annihilation decays $B_s\to \pi\pi$ was firstly calculated in 2004 \cite{Li:2004ep} and the predicted branching fraction was confirmed by LHCb in 2006 \cite{Aaij:2016elb}, which implies that the results of pure annihilation decays based on PQCD approach are reliable.

In this work, we work in the light-cone coordinate. In the $B$ meson rest frame, the momenta of the initial $B$ meson, the final vector meson ($V$) and photon are expressed as
\begin{gather}
P_B=\frac{m_B}{\sqrt{2}}(1,1,\vec{0}_{\perp}),\,\,\,\,\,\nonumber\\
P_V=\frac{m_B}{\sqrt{2}}(r^2,1,\vec{0}_{\perp}),\,\,\,\,\,
P_{\gamma}=\frac{m_B}{\sqrt{2}}(1-r^2,0,\vec{0}_{\perp}),
\end{gather}
with $r=m_V/m_B$, $m_V$ and $m_B$ being the masses of the vector meson and the $B$ meson. For the light quark in the $B$ meson, we denote its momentum as
\begin{eqnarray}
k_1=(x_1\frac{m_B}{\sqrt{2}},0,\mathbf{k}_{1T}),
\end{eqnarray}
where the $x_1$ is the momentum fraction of the light quark and the $\mathbf{k}_{1T}$ is the transverse momentum of the light quark. Similarly, the momentum of the quark of the final vector can be written as
\begin{eqnarray}
k_2&=&(0,x_2\frac{m_B}{\sqrt{2}},\mathbf{k}_{2T}).
\end{eqnarray}
In this work we do not regard the photon as a point-like particle any more and consider its hadronic structure, then the momentum of the light quark in the photon can be written as
\begin{eqnarray}
k_3&=&(\frac{m_B}{\sqrt{2}}(1-r^2)x_3,0,\mathbf{k}_{3T}).
\end{eqnarray}

In PQCD, the decay amplitude of process $B\to M_1M_2$ is factorized into the convolution of the Wilson coefficients (WCs) $C(t)$, the hard kernel $H(x_i,b_i,t)$ and the wave functions of initial and final states, which can be expressed as
\begin{multline}\label{6}
\mathcal{A}=\int_0^1dx_1dx_2dx_3\int_0^{\infty}b_1db_1b_2db_3b_3db_3 C(t)\Phi_B(x_1,b_1) \\
 \Phi_2(x_2,b_2)\Phi_3(x_3,b_3)H(x_i,b_i,t)\exp[-S(x_i,b_i,t)],
\end{multline}
with the $b_i$ being the conjugate variable of the transverse momentum $k_{iT}$. The physics above the scale ($m_b$) has already been absorbed into the WCs $C(t)$. The hard kernel $H(x_i,b_i,t)$ is governed by exchanging one hard gluon, and can be calculated perturbatively. The parameter $t$ is the largest scale appearing in the hard kernel $H(x_i,b_i,t)$. The wave functions describing the soft physics below the factorization scale are not perturbatively calculable but universal, which can be studied in some nonperturbative approaches. The last exponential term is the so-called Sudakov form factor, which arises from the resummation on the double logarithmic terms containing the additional energy scale introduced by the transverse momentum.

In PQCD, the most important inputs are the wave functions, which describe the inner dynamics of the initial and final particles. In the past decades, the wave functions of the $B$ meson and vector mesons have been well studied in the two-body nonleptonic $B$ decays, such as in refs.~ \cite{Keum:2000ph, Lu:2000em, Yu:2005rh,Li:2004ep, Lu:2000hj, Zou:2015iwa, Ali:2007ff}. In general, the wave function $\Phi_{M,\alpha\beta}$ with Dirac indices $\alpha,\beta$ are decomposed into 16 independent components, $1_{\alpha\beta}$, $\gamma^\mu_{\alpha\beta}$, $\sigma^{\mu\nu}_{\alpha\beta}$, $(\gamma^\mu\gamma_5)_{\alpha\beta}$, $\gamma_{5\alpha\beta}$. For the heavy pseudo-scalar $B$ meson, the structure $(\gamma^\mu\gamma_5)_{\alpha\beta}$ and $\gamma_{5\alpha\beta}$ components remain as leading contributions. Therefore, $\Phi_{B,\alpha\beta}$ is written by
\begin{eqnarray} \label{Bwave1}
 \Phi_{B,\alpha\beta} = \frac{i}{\sqrt{2N_c}}\left\{ (\not \! P_B \gamma_5)_{\alpha\beta} \phi_B^A + \gamma_{5\alpha\beta} \phi_B^P
\right\},
\end{eqnarray}
where $N_c = 3$ is the color's degree of freedom, and $\phi_B^{A,P}$ are Lorentz scalar wave functions. According to the heavy quark effective theory, we obtain $\phi_B^P \simeq m_B \phi_B^A$. So,  $B$ meson's wave function can be simplified as
\begin{eqnarray}
\Phi_{B,\alpha\beta}(x,b) = \frac{i}{\sqrt{2N_c}} \left[ (\not \! P_1 \gamma_5)_{\alpha\beta}+ m_B \gamma_{5\alpha\beta}
\right] \phi_B(x,b).
\end{eqnarray}
For the Lorentz scalar wave function $\phi_B(x,b)$, there is a sharp peak at the small $x$ region, we use
\begin{eqnarray}
\phi_B(x,b) = N_B x^2(1-x)^2 \exp \left[-\frac{m_B^2\ x^2}{2 \omega_b^2} -\frac{1}{2} (\omega_b b)^2\right],
\end{eqnarray}
which is adopted in ref. \cite{Keum:2000ph,Lu:2000em}. Noted that $\phi_B$  is normalized by the decay constant $f_B$,
\begin{eqnarray}
 \int_0^1 \!\! dx\  \phi_B (x, b=0)= \frac{f_B}{2\sqrt{2N_c}}.
\end{eqnarray}
As aforementioned, the parameters $\omega_b=0.40\pm0.08$ for $B^0$ meson and $\omega_b=0.50\pm0.10$ for $B^0_s$ meson are almost best fits from the well measured results of the $B_{d,s}\to K\pi$, $\pi \pi$ decays \cite{Keum:2000ph, Lu:2000em, Ali:2007ff}, including the branching fractions and the $CP$ asymmetries.
For the vector mesons $\phi$, $\rho^0$, and $\omega$, we also pursue the same strategy and adopt the same wave functions obtained in QCD sum rules \cite{Ball:2007rt}. Very recently, the Gegenbauer moments in the wave functions of the light mesons have been fitted to available data of branching fractions and direct $CP$ asymmetries globally \cite{Hua:2020usv}, and the fitted results are in good agreement with those in \cite{Ball:2007rt}.

When studying the hard exclusive processes involving photon emission in QCD, a specific feature is that a real photon contains both a pointlike, electromagnetic (EM), and a soft, hadronic component.  Similar to a transversely polarized vector meson, the two-particle distribution amplitudes of a final state photon can be defined as \cite{Ball:2002ps}
\begin{widetext}
\begin{eqnarray}
\langle \gamma(p,\lambda)|\bar
 q(z)\sigma_{\alpha\beta}q(0)|0\rangle
&&=iQ_q\chi(\mu)\langle\bar
qq\rangle(p_\beta\varepsilon_{\gamma\alpha}^*-p_\alpha\varepsilon_{\gamma\beta}^*)\int_0^1
   du e^{iup\cdot z}\phi_\gamma(u,\mu),\,\,\,\,\,\,\,\,\,\,\,\,\,\,\,
 \\
\langle \gamma(p,\lambda)|\bar
 q(z)\gamma_{\alpha}q(0)|0\rangle
&&=-Q_qf_{3\gamma}\varepsilon_{\gamma\alpha}^*\int_0^1
   du e^{iup\cdot z}\psi^{(v)}_\gamma(u,\mu),
 \\
\langle \gamma(p,\lambda)|\bar
 q(z)\gamma_{\alpha}\gamma_5q(0)|0\rangle
&&={\frac{1}{4}}Q_qf_{3\gamma}\epsilon_{\alpha\beta\rho\sigma}p^\rho
z^\sigma\varepsilon_\gamma^{*\beta}\int_0^1
   du e^{iup\cdot z}\psi^{(a)}_\gamma(u,\mu),
\end{eqnarray}
 \end{widetext}
where $\lambda$ and $\varepsilon_\gamma$ denote the polarization and the related polarization vector.  The Lorentz scalar wave functions $\phi_\gamma(u,\mu)$ and $\psi^{(a,v)}_\gamma(u,\mu)$ are twist-2 and twist-3 distribution amplitudes (DAs), respectively. $\langle\bar q q\rangle$ and $\chi(\mu)$ are the quark condensate and the corresponding magnetic susceptibility. $f_{3\gamma}$ is the decay constant of the photon, which appears in the twist-3 DAs. It should be emphasized that $\langle\bar q q\rangle$, $\chi(\mu)$ and $f_{3\gamma}$ are all scale dependence, and their evolution equations can be found in refs.~\cite{Wang:2018wfj, Shen:2018abs}. Finally, we write the momentum space projector for the two-particle LCDAs (up to twist-3) as
\begin{multline} \label{Pwave}
M^\gamma_{\alpha\beta}=\frac{1}{4}Q_q\bigg\{-\langle\bar
qq\rangle(\not\!\varepsilon_\gamma^*\not\!p)\chi(\mu)\phi_\gamma(u,\mu)
-f_{3\gamma}(\not\!\varepsilon^*)\psi^{(v)}_\gamma(u,\mu)\\
-{\frac{1}{8}}f_{3\gamma}\epsilon_{\mu\nu\rho\sigma}(\gamma^{\mu}\gamma^5){\bar n}^\rho
\varepsilon^{*\nu}\Big[n^\sigma{\frac{d}{du}}\psi^{(a)}_\gamma(u,\mu) \\
-2E_\gamma\psi^{(a)}_\gamma(u,\mu){\frac{\partial}{\partial k_{\perp\sigma}}}\Big]\bigg\}_{\alpha\beta},
\end{multline}
and $n=(1,0,{\bf 0}_T)$ and $\bar n=(0,1,{\bf 0}_T)$ are two unit vectors with opposite directions. In our numerical calculations,  the contributions from transverse momentum dependence of the photon wave functions are very small, which indicates that the last part in the third term can be neglected safely.

For the distribution amplitudes $\phi_{\gamma}(u,\mu)$, $\psi^{v}(u,\mu)$ and $\psi^{a}(u,\mu)$, they have been systematically studied in ref.\cite{Ball:2002ps}, and their expressions are written as
\begin{eqnarray}
&&\phi_\gamma(u,\mu) = 6u\bar u  \left[1+\sum_{n=2}^\infty b_n(\mu_0)C^{3/2}_n(\xi)\right],\\
&&\psi^{(v)}(u,\mu)= 10 P_2(\xi)+\frac{15}{8} \left [ 3\omega_{\gamma}^{V}(\mu)- \omega_{\gamma}^{A}(\mu)\right ]P_4(\xi),
\,\,\,\,\, \,\,\,\,\, \\
&&\psi^{(a)}(u,\mu) = \frac{5}{3}C_2^{3/2}(t) \left [ 1 + \frac{9}{16}\omega_{\gamma}^{V}(\mu) -\frac{3}{16}
\omega_{\gamma}^{A}(\mu)  \right ] \,,
 \end{eqnarray}
with $\xi=2u-1$. $C_n^{3/2}(\xi)$ and $P_n(\xi)$ are Gegenbauer polynomials and Legendre polynomials, respectively. Since the scale dependence of the parameters $b_n(\mu)$, $\omega_{\gamma}^{V}(\mu)$ and $\omega_{\gamma}^{A}(\mu)$ have been well studies in refs.\cite{Ball:2002ps, Shen:2018abs}, we will not show them here. Typically, when $\mu=1 {\rm GeV}$, we have
\begin{eqnarray}
b_2=0.07\pm0.07, \,\,\omega_{\gamma}^{V}=3.8\pm1.8,\,\,\omega_{\gamma}^{A}=-2.1\pm1.0.
\end{eqnarray}

Because the massless photon is transverse polarized only, the decay amplitude of the radiative decay $B\to V\gamma$ can only be decomposed into two parts as
\begin{eqnarray}
\mathcal{A}=(\varepsilon_V^*\cdot\varepsilon_{\gamma}^*)\mathcal{A}^S+\frac{1}{P_V\cdot P_{\gamma}}\epsilon_{\mu\nu\rho\sigma}\varepsilon_{\gamma}^{\mu*}\varepsilon_{V}^{\nu*}P_{\gamma}^{\rho}P_V^{\sigma}\mathcal{A}^P,
\end{eqnarray}
with the polarization vectors of the vector meson and the photon $ \varepsilon_V^*$ and $\varepsilon_{\gamma}^*$ respectively. With the calculated $\mathcal{A}^S$ and $\mathcal{A}^P$ in PQCD approach, the branching fraction of the $B\to V\gamma$ decay is
\begin{eqnarray}\label{eq:br}
\mathcal{B}=\tau_B\frac{\mid \mathcal{A}^S\mid^2+\mid\mathcal{A}^P\mid^2}{8\pi m_B}(1-r^2),
\end{eqnarray}
where the $\tau_B$ is the lifetime of the $B$ meson. At the leading order and the leading power,  the decay amplitudes $\mathcal{A}^S$ and $\mathcal{A}^P$  have been calculated in detail in ref. \cite{Li:2006xe}, and we do not present them any more.

\begin{figure*}[!htbp]
\begin{center}
\includegraphics[scale=0.4]{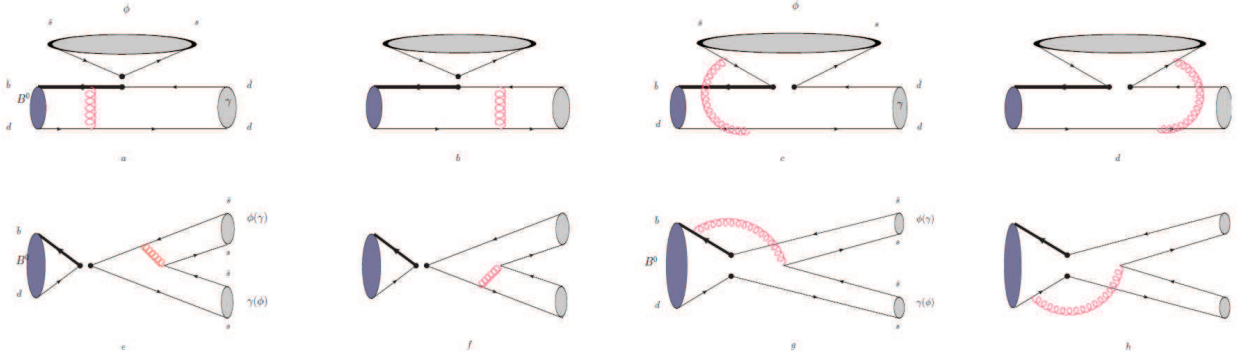}
\caption{The leading order feynman diagrams of $B^0\to \phi \gamma$ decay with the hadronic structure of the photon in PQCD.}
\label{fig-1}
\end{center}
\end{figure*}

Now, we come to study the contributions from the HSP to decays $B\to\phi\gamma$ and $B_s\to(\rho^0,\omega)\gamma$. In the following work, we will take $B^0\to \phi\gamma$ as an example for illustration, which is governed by the $b\to d \bar s s$ transition. According to the effective Hamiltonian \cite{Buchalla:1995vs}, the corresponding diagrams are plotted in Fig. \ref{fig-1}. The decay amplitude with the contribution of HSP can be expressed as
\begin{eqnarray}\label{amplitudes}
\mathcal{A}^{i}_{H}(B\to\phi\gamma)&=&-V_{td}V^*_{tb}\bigg[\left(a_3+a_5-\frac{a_7}{2}-\frac{a_9}{2}\right)\mathcal{A}^{i(LL)}_{H,ef}\nonumber\\
&&+\left(C_4-\frac{C_{10}}{2}\right)\mathcal{A}^{i(LL)}_{H,enf}\nonumber\\
&&+\left(C_6-\frac{C_{8}}{2}\right)\mathcal{A}^{i(SP)}_{H,enf}\nonumber\\
&&+\left(a_3-\frac{a_9}{2}\right)\left(\mathcal{A}^{i(LL)}_{H,af1}+\mathcal{A}^{i(LL)}_{H,af2}\right)\nonumber\\
&&+\left(a_5-\frac{a_7}{2}\right)\left(\mathcal{A}^{i(LR)}_{H,af1}+\mathcal{A}^{i(LR)}_{H,af2}\right)\nonumber\\
&&+\left(C_4-\frac{C_{10}}{2}\right)\left(\mathcal{A}^{i(LL)}_{H,anf1}+\mathcal{A}^{i(LL)}_{H,anf2}\right)\nonumber\\
&&+\left(C_6-\frac{C_{8}}{2}\right)\left(\mathcal{A}^{i(SP)}_{H,anf1}+\mathcal{A}^{i(SP)}_{H,anf2}\right)\bigg],
\end{eqnarray}
with $i=S,P$. The combined coefficients $a_i$ are defined as
\begin{eqnarray}
a_3=C_3+C_4/3,&\;\;& a_5=C_5+C_6/3,\nonumber\\
a_7=C_7+C_8/3,&\;\;& a_9=C_9+C{10}/3.
\end{eqnarray}
The subscripts ``$ef$" and ``$enf$" stand for the factorizable and nonfactorizable diagrams with the emission of $\phi$ meson, respectively. Similarly, the ``$af1(2)$" and ``$anf1(2)$" means the factorizable and nonfactorizable annihilation diagrams, where the number ``1(2)" means the case that the produced strange quark enters into the $\phi$ meson ($\gamma$), respectively. The superscripts ``$LL$", ``$LR$" and ``$SP$" represent the inserted $(V-A)\otimes(V-A)$, $(V-A)\otimes(V+A)$, and $(S-P)\otimes(S+P)$ currents. The expressions of the diagrams with different operators are given in the Appendix. Combining the eqs.(\ref{eq:br}) and (\ref{amplitudes}), we then calculate the branching fraction of this decay.

\section{Results and Discussions}\label{sec:results}
In this section, we first list  some parameters in our numerical calculations as \cite{Zyla:2020zbs,Ball:2007rt}
\begin{gather}
f_{B} = 210\pm20 \mbox{ MeV},\
f_{B_s} = 230\pm20\mbox{ MeV},\nonumber
\\
f_{\phi}^T = 186\pm9 \mbox{ MeV},\
f_{\rho}^T = 165\pm9 \mbox{ MeV},
f_{\omega}^T = 151\pm9 \mbox{ MeV},\nonumber
\\
V_{tb}=0.999172_{0.000035}^{+0.000024},\
V_{ts}=0.03978_{-0.00060}^{+0.00082},\nonumber\\
V_{td}=0.00854_{-0.00016}^{+0.00023}, \nonumber
\\
\tau_{B}=1.519\times10^{-12}s,\
\tau_{B_s}=1.527\times10^{-12}s.
\end{gather}
Using the above input parameters, we then calculate the branching fractions of $B^0 \to \phi \gamma$,$B_s \to \rho^0\gamma$ and $B_s \to \omega\gamma$ including corrections from HSP at leading order, and list the results in Table.~\ref{br}. In addition, the results of the leading power are also presented for a comparison. In our calculations, three kinds of errors have been studied, which are from uncertainties of the nonperturbative physics, the unknown QCD corrections and the CKM matrix parameters, respectively. The first errors are from the uncertainties of the parameters in the wave functions, such as the decays constants, the inner parameters in the distribution amplitudes of the initial and the final mesons. It should be emphasized that this kind of errors are dominant, and more precise results from nonperturbative approaches are called in future. The second uncertainties come from the QCD scale $\Lambda_{QCD}$ and the hard kernel scale $t$, whose variants reflect the effects of the higher order QCD corrections. In this work, we set $\Lambda_{QCD}=(0.25\pm0.05){\rm GeV}$ and vary $t$ from $0.8t$ to $1.2t$. The last errors are caused by the CKM matrix elements.

\begin{table}[!tbh]
\caption{The leading order (LO) branching ratios and the improved branching ratios (IBR) of the pure annihilation type $B\to V\gamma$ decays in PQCD approach.} \label{br}
\begin{center}
\begin{tabular}{l c c}
 \hline \hline
 \multicolumn{1}{c}{Decay Modes}&\multicolumn{1}{c}{LO($10^{-12}$)} &\multicolumn{1}{c}{IBR($10^{-11}$)} \\
\hline\hline
 $B^0 \to \phi    \gamma$       &$0.9^{+0.2+0.3+0.0}_{-0.2-0.4-0.0}$        &$3.57^{+2.12+0.25+0.19}_{-1.58-0.31-0.14}$ \\
 $B_s \to \rho^0 \gamma$       &$72.8^{+26.5+4.6+2.4}_{-21.0-4.4-3.3}$    &$12.4_{-3.87-0.55-0.60}^{+4.98+1.04+0.70}$ \\
 $B_s \to \omega\gamma$       &$8.6^{+5.3+6.9+0.2}_{-3.4-4.2-0.2}$         &$35.1_{-21.2-4.14-1.90}^{+31.8+2.13+1.70}$  \\
 \hline \hline
\end{tabular}
\end{center}
\end{table}

From Table.~\ref{br}, one finds that at the leading order the branching fractions of the $B_s\to (\rho^0,\omega)\gamma$ decays are much larger than that of $B^0\to\phi\gamma$, it is because the former decays have the $\bar uu$ component and thus get the contributions from the tree operators with larger WCs. It is important to note that when the contributions from HSP are included, the branching fractions of all concerned decays are enhanced remarkably, and the branching fraction of $B\to \phi\gamma$ increases by about even  40 times compared with that of leading power. At leading power,  $B\to \phi\gamma$ is governed by the transition $b\to d \bar s s$, and the produced $\bar s$ and $ s$ quarks form the final $\phi$ meson. The photon can be radiated from any quark participating in the weak interaction. Due to the symmetry, the contributions of the photon radiated from the strange and anti-strange quarks are cancelled by each other, therefore only the diagrams with photons emitted from the beauty and down quarks contribute to the amplitudes. The explicit amplitudes are given in ref.~\cite{Li:2006xe}. When the contributions of HSP are included, more diagrams contribute to the amplitudes $\mathcal{A}^S$ and $\mathcal{A}^P$, as shown in Fig.1. Compared with diagrams of leading power, both factorizable and nonfactorizable emission diagrams contribute to the amplitudes without cancellations. Furthermore, for the nanfactorizable diagrams, besides the $(V-A)(V-A)$ operators, the $(S-P)(S+P)$ operator which results from the Fierz transformation of the $(V-A)(V+A)$ also takes large contributions. The picture is very similar to the $B \to \rho^0\phi$ decay. After summing all contributions, we find that the amplitude from HSP is much larger than that of leading power, leading to a larger branching fraction.

Generally, the power expansion ($1/m_B$) in $B$ meson decays is an effective expansion in most of decay modes such as the two-body non-leptonic decays. As stated in Sec.\ref{sec:introduction}, there are many sources of the power corrections, such as the high-twist DAs of $B$ and light mesons, the soft corrections to the leading twist contributions and the subleading power “hadronic” photon correction. For the power corrections from $B$ and light mesons, their contributions are less than $20\%$. Such corrections would be plagued by large theoretical uncertainties, and we have not discussed these effects yet in current work. However, for the pure annihilation radiative $B\to V\gamma$ decays, the leading power contributions (annihilation diagrams) are already power suppressed compared with decays dominated by the emission type diagrams. When HSP is taken into account, the emission diagrams are involved, which will enhance the next-to-leading power contributions and make the higher power corrections significantly larger than that of the leading power. All these can be neatly summarized in a short conclusion: for some special decay modes, if the leading power contributions are suppressed by some mechanism, the higher power contributions could be larger than that of the leading power, which leads to that the branching fraction might be enhanced significantly.

We also note that the branching fractions of $B_s \to \rho^0\gamma$ and $B_s \to \omega\gamma$ are at the order of ${\cal O}(10^{-10})$, and that of $B \to \phi\gamma$ is at the order of ${\cal O}(10^{-11})$. Unfortunately, such small branching fractions cannot be measured in the current experiments. In order to highlight the effects of the corrections from HSP, we then define a ratio as
\begin{eqnarray}
\mathcal{R}_{\rho\omega}=\frac{\mathcal{B}(B_s\to\rho^0\gamma)}{\mathcal{B}(B_s\to\omega\gamma)},\;\;
\end{eqnarray}
which can reduce of the dependence on the nonperturbative parameters effectively. At leading power, $\mathcal{R}_{\rho\omega}\sim 8$ is much larger than 1, while this ratio becomes 0.35 when HSP are involved. It is apparent that this ratio decreases about 40 times, thus it may be a good probe for testing the effects of HSP when the data are available. At leading order, because of the interference between the $u\bar{u}$ and $d\bar{d}$ components, $B_s \to \rho^0\gamma$ has larger penguin contributions than $B_s \to \omega\gamma$, leading to that the branching fractions of $B_s \to \rho^0\gamma$ is much larger than that of  $B_s \to \omega\gamma$. When including the HSP corrections, although they are suppressed by the CKM matrix elements and the power, the new introduced tree contributions by $\bar u u$ component in photon are comparable with those of penguin contributions, due to the large WCs, especially in the nonfactorizable diagrams. Because the signs of $\bar dd$ in $\rho^0$ and $\omega$ are different, the interferences  between contributions of tree operators and penguins ones are different for $B_s \to \rho^0\gamma$ and $B_s \to \omega\gamma$ decays. Again, due to the interference between tree and the penguin contributions, the HSP corrections to $B_s\to\rho^0\gamma$ are much smaller than that to $B_s\to\omega\gamma$, leading to a small $R_{\rho\omega}$.

As aforementioned, the HSP corrections could increase the branching fractions of these pure annihilation decays remarkably. In ref. \cite{Shen:2018abs}, it is found that the contribution of HSP plays important role in decay $B\to\gamma l \nu$ and the branching faction can be decreased by $20\%$.  In ref.~\cite{Lu:2006nza}, the contributions of electromagnetic penguin operators to $B_s\to (\rho^0,\omega)\gamma$ decays have been also examined in QCDF, and the results indicated that they are suppressed by the electromagnetic coupling constant $\alpha_e$. Very recently, these three decays have been comprehensively analyzed at leading power in the framework of soft-collinear effective theory in \cite{Deng:2021zoi}, where the authors found that the $\phi-\omega-\rho^0$ mixing effect could enhance the branching fractions of $B^0 \to \phi\gamma$ and $B_s\to \omega \gamma$  about 3 orders of magnitude. If their conclusion holds in PQCD, the branching fractions of $B^0 \to \phi\gamma$ and $B_s\to \omega\gamma$ are roughly estimated to be at the order of $10^{-8}$ and $10^{-7}$, respectively, which might be measurable in Belle-II and LHCb experiments. The future measurements on these decays are helpful to discern all the above physics mechanisms.
\section{Summary}\label{sec:summay}
As the pure annihilation type radiative $B$ decays suppressed by $\Lambda/m_B$, the decays $B\to\phi\gamma$ and $B_s\to(\rho^0,\omega)\gamma$ governed by the flavor changing neutral currents have small branching fractions, which make them very sensitive to the effects of the new physics beyond the standard model. Before studying the effects of new physics, it is necessary for us to calculated the observables in SM with high precision. In this work, we mainly investigated the power corrections from the hadronic structure of photon in these pure annihilation decays. Because the quark components in photon are taken in account, the emission diagrams are involved, which are not suppressed in comparing to the annihilation diagrams. Thus, the branching fraction of $B\to \phi\gamma$ is enhanced by about 40 times, and the branching fractions of $B_s \to \rho^0\gamma$ and $B_s \to \omega\gamma$ are at the order of ${\cal O}(10^{-10})$. Moreover, in order to shed light on this kind of corrections and to reduce the theoretical uncertainties, we also define a ratio of branching fractions of $B_s \to \rho^0\gamma$ and $B_s \to \omega\gamma$, namely $\mathcal{R}_{\rho\omega}$, and find that this ratio is changed significantly by hadronic structure of photon. All above results could be tested in the high luminosity accelerators in future.

\section*{Acknowledgment}
This work was supported in part by the Open Project of Shaanxi Collaborative Innovation Center of Industrial Auxiliary Chemistry $\&$ Technology (No. XTKF-2020-01),  by the National Science Foundation of China under the Grant Nos. 11705159 and 11975195, and the Natural Science Foundation of Shandong province under the Grant Nos.ZR2019JQ04 and ZR2020MA093, by the Project of Shandong Province Higher Educational Science and Technology Program under Grants No. 2019KJJ007.

\begin{widetext}
\begin{appendix}
\section{Amplitudes}
The amplitudes of factorizable emission diagrams ($a$ and $b$) with the $(V-A)(V-A)$ current are given as
\begin{multline}
\mathcal{A}^{S(LL)}_{H,ef}=\sqrt{\frac{3}{2}}\pi C_Ff_{\phi}^T r_V Q_d m_B^3V_{tb}V_{td}^*\int_0^1dx_1dx_3\int_0^{\infty}b_1db_1b_3db_3\phi_B(x_1,b_1)\\
\Bigg\{\Big[f_{3\gamma}\Big(x_3\phi_{\gamma}^a(x_3)+4(x_3+2)\phi_{\gamma}^v(x_3)\Big)
-4m_B\chi(\mu)\langle \bar{d}d\rangle\phi_{\gamma}(x_3)\Big]C(t_a)E_{ef}(t_a)h_{ef}[x_1,x_3(1-r_V^2),b_1,b_3]\\
-f_{3\gamma}\Big[\phi_{\gamma}^a(x_3)-4\phi_{\gamma}^v(x_3)\Big]C(t_b)E_{ef}(t_b)h_{ef}[x_3,x_1(1-r_V^2),b_3,b_1]\Bigg\};
\end{multline}
\begin{multline}
\mathcal{A}^{P(LL)}_{H,ef}=\sqrt{\frac{3}{2}}\pi C_F f_{\phi}^T r_V Q_d m_B^3V_{tb}V_{td}^*\int_0^1dx_1dx_3\int_0^{\infty}b_1db_1b_3db_3\phi_B(x_1,b_1)\\
\Bigg\{\Big[f_{3\gamma}\Big((2+x_3)\phi_{\gamma}^a(x_3)+4x_3\phi_{\gamma}^v(x_3)\Big)
+4m_B\chi(\mu)\langle\bar{d}d\rangle\phi_{\gamma}(x_3)\Big]C(t_a)E_{ef}(t_a)h_{ef}(x_1,x_3(1-r_V^2),b_1,b_3)\\
+f_{3\gamma}\Big[\phi_{\gamma}^a(x_3)-4\phi_{\gamma}^v(x_3)\Big]C(t_b)E_{ef}(t_b)h_{ef}[x_3,x_1(1-r_V^2),b_3,b_1]\Bigg\},
\end{multline}
with the color factor $C_F=4/3$ and the $\phi(1020)$ meson decay constant $f_{\phi}^T$. $Q_d$ is the charge of the down quark in the hadronic structure of the photon. The Sudakov form factors $E_{ef}$, hard functions$h_{ef}$ and the scales $t_{a,b}$ can be found in ref.\cite{Zou:2015iwa}, because the behaviors of the photon are very similar to a vector.

For the nonfactorizable emission diagrams ($c$ and $d$), not only $(V-A)(V-A)$ current but also $(S-P)(S+P)$ current can be inserted, and the corresponding amplitudes are written as
\begin{multline}
\mathcal{A}^{S(LL)}_{H,enf}=\pi C_Fr_Vm_B^3Q_d\int_0^1dx_1dx_2dx_3\int_0^{\infty}b_1db_1b_2db_2 \phi_B(x_1,b_1) \\
\Bigg\{\left[8(x_2-1)\phi_{\gamma}(x_3)\chi(\mu)m_B\langle\bar{d}d\rangle\Big(\phi_{\phi}^a(x_2)+\phi_{\phi}^v(x_2)\Big)\right]
E_{enf}(t_a)h_{enfa}[x_1,x_2,x_3,b_1,b_2] \\
-4\left[2 x_2\phi_{\gamma}(x_3)\chi(\mu)m_B\langle\bar{d}d\rangle\Big(\phi_{\phi}^a(x_2)+\phi_{\phi}^v(x_2)\Big)
-f_{3\gamma}(x_2+x_3)\Big(\phi_{\gamma}^a(x_3)\phi_{\phi}^a(x_2)-4\phi_{\gamma}^v(x_3)\phi_{\phi}^v(x_2)\Big)   \right] \\
   E_{enf}(t_b)h_{enfb}[x_1,x_2,x_3,b_1,b_2]\Bigg\};
\end{multline}
\begin{multline}
\mathcal{A}^{P(LL)}_{H,enf}=\pi C_Fr_Vm_B^3Q_d\int_0^1dx_1dx_2dx_3\int_0^{\infty}b_1db_1b_2db_2
\phi_B(x_1,b_1) \\
\Bigg\{\left[8(1-x_2)\phi_{\gamma}(x_3)\chi(\mu)m_B\langle\bar{d}d\rangle\Big(\phi_{\phi}^a(x_2)+\phi_{\phi}^v(x_2)\Big)\right]
E_{enf}(t_a)h_{enfa}[x_1,x_2,x_3,b_1,b_2] \\
+4\left[2 x_2\phi_{\gamma}(x_3)\chi(\mu)m_B\langle\bar{d}d\rangle\Big(\phi_{\phi}^a(x_2)+\phi_{\phi}^v(x_2)\Big)
+f_{3\gamma}(x_2+x_3)\Big(4\phi_{\gamma}^v(x_3)\phi_{\phi}^v(x_2)-\phi_{\gamma}^a(x_3)\phi_{\phi}^a(x_2)\Big)   \right] \\
   E_{enf}(t_b)h_{enfb}[x_1,x_2,x_3,b_1,b_2]\Bigg\};
\end{multline}
\begin{multline}
\mathcal{A}^{S(SP)}_{H,enf}=\pi C_Fr_Vm_B^3Q_d\int_0^1dx_1dx_2dx_3\int_0^{\infty}b_1db_1b_2db_2
\phi_B(x_1,b_1) \\
\Bigg\{4\left[2(x_2-1)\phi_{\gamma}(x_3)\chi(\mu)m_B\langle\bar{d}d\rangle \Big(\phi_{\phi}^a(x_2)-\phi_{\phi}^v(x_2)\Big)
-f_{3\gamma}(x_2-x_3-1)\Big(4\phi_{\gamma}^v(x_3)\phi_{\phi}^v(x_2)+\phi_{\gamma}^a(x_3)\phi_{\phi}^a(x_2)\Big)   \right] \\
 \times  E_{enf}(t_a)h_{enfa}[x_1,x_2,x_3,b_1,b_2]
+\left[8 x_2\phi_{\gamma}(x_3)\chi(\mu)m_B\langle\bar{d}d\rangle\Big(\phi_{\phi}^v(x_2)-\phi_{\phi}^a(x_2)\Big)\right]  E_{enf}(t_b)h_{enfb}[x_1,x_2,x_3,b_1,b_2]\Bigg\};
\end{multline}
\begin{multline}
\mathcal{A}^{P(SP)}_{H,enf}=\pi C_Fr_Vm_B^3Q_d\int_0^1dx_1dx_2dx_3\int_0^{\infty}b_1db_1b_2db_2\phi_B(x_1,b_1) \\
\Bigg\{4\left[2(1-x_2)\phi_{\gamma}(x_3)\chi(\mu)m_B\langle\bar{d}d\rangle \Big(\phi_{\phi}^a(x_2)-\phi_{\phi}^v(x_2)\Big)
+f_{3\gamma}(1-x_2+x_3)\Big(4\phi_{\gamma}^v(x_3)\phi_{\phi}^v(x_2)+\phi_{\gamma}^a(x_3)\phi_{\phi}^a(x_2)\Big)   \right] \\
 \times  E_{enf}(t_a)h_{enfa}[x_1,x_2,x_3,b_1,b_2]
+\left[8 x_2\phi_{\gamma}(x_3)\chi(\mu)m_B\langle\bar{d}d\rangle\Big(\phi_{\phi}^a(x_2)-\phi_{\phi}^v(x_2)\Big)\right]  E_{enf}(t_b)h_{enfb}[x_1,x_2,x_3,b_1,b_2]\Bigg\}.
\end{multline}

For the annihilation diagrams ($e$-$h$), because the produced strange quark can enter not only into the $\phi$ meson, but also into the photon, we have to discuss two cases. If the strange quark enters into the photon, the corresponding factorizable and nonfactorizable annihilation amplitudes with possible currents are given as
\begin{multline}
\mathcal{A}^{S(LL)}_{H,af1}=-\sqrt{\frac{3}{2}}\pi C_F f_b Q_s r_V f_{3\gamma}m_B^3\int_0^1dx_2dx_3\int_0^{\infty}b_2db_2b_3db_3  \\
\Bigg\{\left[x_3\Big(\phi_{\gamma}^a(x_3)+4\phi_{\gamma}^v(x_3)\Big)\Big(\phi_{\phi}^a(x_2)-\phi_{\phi}^v(x_2)\Big)
-2\phi_{\gamma}^a(x_3)\phi_{\phi}^a(x_2)+8\phi_{\gamma}^v(x_3)\phi_{\phi}^v(x_2)\right]
   E_{af}(t_e)h_{afe}[x_2,x_3,b_2,b_3] \\
+\left[x_2\Big(\phi_{\gamma}^a(x_3)-4\phi_{\gamma}^v(x_3)\Big)\Big(\phi_{\phi}^a(x_2)+\phi_{\phi}^v(x_2)\Big)
+\Big(\phi_{\gamma}^a(x_3)+4\phi_{\gamma}^v(x_3)\Big)\Big(\phi_{\phi}^a(x_2)-\phi_{\phi}^v(x_2)\Big)\right]
 E_{af}(t_f)h_{aff}[x_2,x_3,b_2,b_3]\Bigg\};
\end{multline}
\begin{multline}
\mathcal{A}^{P(LL)}_{H,af1}=\sqrt{\frac{3}{2}}\pi C_F f_b Q_s r_V f_{3\gamma}m_B^3\int_0^1dx_2dx_3\int_0^{\infty}b_2db_2b_3db_3 \\
\Bigg\{\left[\phi_{\gamma}^a(x_3)\Big((x_3-2)\phi_{\phi}^v(x_2)-x_3\phi_{\phi}^a(x_2)\Big)
-4\phi_{\gamma}^v(x_3)\Big((x_3-2)\phi_{\phi}^a(x_2)-x_3\phi_{\phi}^v(x_2)\Big)\right]E_{af}(t_e)h_{afe}[x_2,x_3,b_2,b_3] \\
+\left[x_2\Big(\phi_{\gamma}^a(x_3)-4\phi_{\gamma}^v(x_3)\Big)\Big(\phi_{\phi}^a(x_2)+\phi_{\phi}^v(x_2)\Big)-\Big(\phi_{\gamma}^a(x_3)
+4\phi_{\gamma}^v(x_3)\Big)\Big(\phi_{\phi}^a(x_2)-\phi_{\phi}^v(x_2)\Big)\right]
E_{af}(t_f)h_{aff}[x_2,x_3,b_2,b_3]\Bigg\};
\end{multline}
\begin{eqnarray}
\mathcal{A}^{S(LR)}_{H,af1}=\mathcal{A}^{S(LL)}_{H,af1},
\end{eqnarray}
\begin{eqnarray}
\mathcal{A}^{P(LR)}_{H,af1}=-\mathcal{A}^{P(LL)}_{H,af1};
\end{eqnarray}
\begin{multline}
\mathcal{A}^{S(LL)}_{H,anf1}=4\pi C_F Q_s r_V m_B^3\int_0^1dx_1dx_2dx_3\int_0^{\infty}b_1db_1b_2db_2 \phi_B(x_1,b_1)\\
\Bigg\{\left[f_{3\gamma}\Big(\phi_{\gamma}^a(x_3)\phi_{\phi}^a(x_2)-4\phi_{\gamma}^v(x_3)\phi_{\phi}^v(x_2)\Big)
+2(x_2-1)r_V\phi_{\gamma}(x_3)\phi_{\phi}^t(x_2)\chi(\mu)m_B\langle\bar{s}s\rangle\right] E_{anf}(t_g)h_{anfg}[x_1,x_2,x_3,b_1,b_2]\\
-2\left[x_2 r_V\phi_{\gamma}(x_3)\phi_{\phi}^t(x_2)\chi(\mu)m_B\right]E_{anf}(t_h)h_{anfh}[x_1,x_2,x_3,b_1,b_2]\Bigg\};
\end{multline}
\begin{multline}
\mathcal{A}^{P(LL)}_{H,anf1}=4\pi C_F Q_s r_V m_B^3\int_0^1dx_1dx_2dx_3\int_0^{\infty}b_1db_1b_2db_2 \phi_B(x_1,b_1)\\
\Bigg\{\left[f_{3\gamma}\Big(4\phi_{\gamma}^v(x_3)\phi_{\phi}^a(x_2)-\phi_{\gamma}^a(x_3)\phi_{\phi}^v(x_2)\Big)
+2(x_2-1)r_V\phi_{\gamma}(x_3)\phi_{\phi}^t(x_2)\chi(\mu)m_B\langle\bar{s}s\rangle\right]E_{anf}(t_g)h_{anfg}[x_1,x_2,x_3,b_1,b_2]\\
-2\left[x_2 r_V\phi_{\gamma}(x_3)\phi_{\phi}^t(x_2)\chi(\mu)m_B\right]E_{anf}(t_h)h_{anfh}[x_1,x_2,x_3,b_1,b_2]\Bigg\};
\end{multline}
\begin{eqnarray}
\mathcal{A}^{S(SP)}_{H,anf1}=\mathcal{A}^{S(LL)}_{H,anf1},
\end{eqnarray}
\begin{eqnarray}
\mathcal{A}^{P(SP)}_{H,anf1}=-\mathcal{A}^{P(LL)}_{H,anf1}.
\end{eqnarray}

When the strange quark enters into the $\phi$ meson, the amplitudes are expressed as
\begin{eqnarray}
\mathcal{A}^{S(LL)}_{H,af2}
=\mathcal{A}^{S(LR)}_{H,af2}
=\mathcal{A}^{S(LL)}_{H,af1},
\end{eqnarray}
\begin{eqnarray}
\mathcal{A}^{P(LL)}_{H,af2}
=-\mathcal{A}^{P(LR)}_{H,af2}
=-\mathcal{A}^{P(LL)}_{H,af2},
\end{eqnarray}
\begin{multline}
\mathcal{A}^{S(LL)}_{H,anf2}=4\pi C_F Q_s r_V m_B^3\int_0^1dx_1dx_2dx_3\int_0^{\infty}b_1db_1b_2db_2 \phi_B(x_1,b_1) \\
\Bigg\{\left[f_{3\gamma}\Big(\phi_{\gamma}^a(x_2)\phi_{\phi}^a(x_3)-4\phi_{\gamma}^v(x_2)\phi_{\phi}^v(x_3)\Big)
-2x_3 r_V\phi_{\gamma}(x_2)\phi_{\phi}^t(x_3)\chi(\mu)m_B\langle\bar{s}s\rangle\right] E_{anf}(t_g)h_{anfg}[x_1,x_2,x_3,b_1,b_2] \\
+2\left[(x_3-1)r_V\phi_{\gamma}(x_2)\phi_{\phi}^t(x_3)\chi(\mu)m_B\right]
E_{anf}(t_h)h_{anfh}[x_1,x_2,x_3,b_1,b_2]\Bigg\};
\end{multline}
\begin{multline}
\mathcal{A}^{P(LL)}_{H,anf2}=4\pi C_F Q_s r_V m_B^3\int_0^1dx_1dx_2dx_3\int_0^{\infty}b_1db_1b_2db_2 \phi_B(x_1,b_1) \\
\Bigg\{\left[f_{3\gamma}\Big(4\phi_{\gamma}^v(x_2)\phi_{\phi}^a(x_3)-4\phi_{\gamma}^a(x_2)\phi_{\phi}^v(x_3)\Big)
+2x_3 r_V\phi_{\gamma}(x_2)\phi_{\phi}^t(x_3)\chi(\mu)m_B\langle\bar{s}s\rangle\right]E_{anf}(t_g)h_{anfg}[x_1,x_2,x_3,b_1,b_2] \\
-2\left[(x_3-1)r_V\phi_{\gamma}(x_2)\phi_{\phi}^t(x_3)\chi(\mu)m_B\right] E_{anf}(t_h)h_{anfh}[x_1,x_2,x_3,b_1,b_2]\Bigg\};
\end{multline}
\begin{eqnarray}
\mathcal{A}^{S(SP)}_{H,anf2}
=\mathcal{A}^{S(LL)}_{H,anf2},
\end{eqnarray}
\begin{eqnarray}
\mathcal{A}^{P(SP)}_{H,anf2}
=-\mathcal{A}^{P(LL)}_{H,anf2}.
\end{eqnarray}
\end{appendix}

\end{widetext}


\end{document}